\documentclass[aps,prl,twocolumn,superscriptaddress,amsmath,amssymb,noeprint]{revtex4-2}

\usepackage{hyperref} 
\usepackage{graphicx}
\usepackage{float}
\usepackage{amsmath}
\usepackage{mathrsfs}
\usepackage{soul}
\usepackage{amssymb}
\usepackage{booktabs}
\usepackage{siunitx}
\usepackage{braket}
\usepackage{upgreek}
\usepackage{bm}
\usepackage{multirow}
\usepackage{xcolor}
\usepackage[normalem]{ulem}
\usepackage{setspace}
\usepackage{amsfonts}
\usepackage{stackengine}

\DeclareGraphicsExtensions{.png,.jpg,.eps}

\begin{document}

\title{Comment on ``Observation of Kardar--Parisi--Zhang universal scaling in two dimensions''}

\author{J.~Bloch}
\affiliation{Universit\'{e} Paris-Saclay, CNRS, Centre de Nanosciences et de Nanotechnologies (C2N), 91120, Palaiseau, France}

\author{M. Escalera}
\affiliation{Universit\'{e} Paris-Saclay, CNRS, Centre de Nanosciences et de Nanotechnologies (C2N), 91120, Palaiseau, France}

\author{Q. Fontaine}
\affiliation{Universit\'{e} Paris-Saclay, CNRS, Centre de Nanosciences et de Nanotechnologies (C2N), 91120, Palaiseau, France}

\author{F. Helluin}
\affiliation{Universit\'{e} Paris-Saclay, CNRS, Centre de Nanosciences et de Nanotechnologies (C2N), 91120, Palaiseau, France}

\author{A. Minguzzi}
\affiliation{Univ. Grenoble Alpes and CNRS, Laboratoire de Physique et Mod\'elisation des Milieux Condens\'es, 38000 Grenoble, France.}

\author{L. Canet}
\affiliation{Univ. Grenoble Alpes and CNRS, Laboratoire de Physique et Mod\'elisation des Milieux Condens\'es, 38000 Grenoble, France.}

\author{S. Ravets}
\affiliation{Universit\'{e} Paris-Saclay, CNRS, Centre de Nanosciences et de Nanotechnologies (C2N), 91120, Palaiseau, France}

\date{\today}

\begin{abstract}
In their paper published in Science {\bf 392}, 221 (2026), Widmann and collaborators reported interferometry experiments to explore the emission coherence decay of a two-dimensional polariton condensate generated in an array of coupled resonators. The authors claim evidence of Kardar--Parisi--Zhang (KPZ) universal scaling  in the measured spatio-temporal coherence decay. 
We argue in the following that the data were not properly analyzed. 
We  re-analyze the experimental data  acquired both with the square and the triangular lattices for  various values of the 
excitation power.
Instead of stretched exponential decays, in the space (time) windows considered in the paper we find that the measured $|g^{(1)}(\delta {\bf r}, \delta t)|$ at $\delta t=0$ ($|\delta {\bf r}| $ close to $0$) rather show Gaussian (exponential) decay  for all excitation powers. As a result, 
 using as temporal and spatial exponents $\beta=0.5$ and $\chi=1$, the data for all pump powers are found to collapse
onto a single curve, which is not the KPZ scaling function.  In particular, we show that the data collapse onto the KPZ scaling function presented in the paper is an artifact stemming from incorrect data normalization.
We thus conclude that the main claim of the paper is not justified as the spatio-temporal decays of the coherence over the space-time windows analyzed in the paper are not well described by the KPZ universal behavior.
\end{abstract}

\maketitle 

\onecolumngrid

In this comment of the analysis presented in Ref.~\cite{Widmann2026}, we first briefly review in Sec.~\ref{sec:theor} the main theoretical results related to the space-time decay of the first-order coherence of driven-dissipative exciton-polariton condensates. We then present in Secs.~\ref{sec:scaling} and \ref{sec:collapse} our analysis of the data acquired in the experiment reported in Ref.~\cite{Widmann2026}, which  robustly shows that the decay of the coherence is exponential in time and Gaussian in space for all pump powers and both the square and triangular lattice geometries, in contrast to the claim of this paper. We discuss in particular the collapse of the data in Sec.~\ref{sec:norm}, highlighting the importance of the normalization, and showing that, without it, such a collapse can yield misleading results. For completeness, we provide in  the Appendix the re-analysis of all the data available with Ref.~\cite{Widmann2026} and its Supplemental Material, presenting a systematic comparison of the collapse obtained with proper normalization using either the KPZ exponents or the exponential-Gaussian ones, which demonstrates that the latter is 
the most relevant modeling for all pump powers and both lattice geometries.

\section{Theoretical background}
\label{sec:theor}

\subsection{Space-time decay of the first-order coherence}
In 2015, it was demonstrated that, under certain assumptions,  the behavior of the first-order correlation function
of a driven-dissipative condensate is completely dominated 
by the dynamics of the phase of the condensate
 which grows indefinitely, while its density fluctuations are negligible  since they remain bounded \cite{Altman2015}. 
It follows that, in this regime
\begin{equation}
    g^{(1)}(\delta \boldsymbol{r}, \delta t) \propto e^{-\frac{1}{2}C_{\theta\theta}(\delta \boldsymbol{r}, \delta t)}  \,,
    \label{eq:g1}
\end{equation}
where 
$C_{\theta\theta}(\delta \boldsymbol{r}, \delta t) = \langle \left[ \theta(\delta \boldsymbol{r}, \delta t) - \theta(\boldsymbol{0},0)\right]^2 \rangle$ is the two-point correlation function of the phase. Moreover, it was also shown \cite{Altman2015} that  the phase dynamics can be mapped onto the Kardar--Parisi--Zhang equation, which implies that $C_{\theta\theta}$ is expected to follow the KPZ universal behavior, which is given by the scaling form
\begin{equation}
C_{\theta\theta}(\delta \boldsymbol{r}, \delta t) = C_0 \, \delta t^{2\beta}F(C_1|\delta\boldsymbol{r}|/\delta t^{1/z}) \sim \left\{\begin{array}{l l}
  A \, \delta t^{2\beta} \qquad&  {\rm for} \,\,|\delta \boldsymbol{r}|=0\\
 B\,|\delta \boldsymbol{r}|^{2\chi} &  {\rm for} \,\, \delta t=0
\end{array}\right. \,,
\label{KPZ_scaling}
\end{equation}
where $\beta$, $\chi$ and $z$ are universal exponents and $F(y)$ is a universal scaling function. The dynamical exponent $z$ is defined according to $z = \chi/\beta$, $C_0$, $C_1$ are non-universal constants, and $A$, $B$ are two amplitudes related to these constants and to the microscopic parameters entering the KPZ equation. In dimension $d=2$, the values of the KPZ critical exponents have been estimated numerically and are approximately given by $\beta\approx0.24$, $\chi\approx0.39$, $z\approx1.62$ \cite{Pagnani2015}. The  KPZ universal scaling function $F(y)$ has been calculated in $d=2$ within the functional renormalization group approach \cite{Kloss2012}.\\

By combining Eqs.(\ref{eq:g1}) and (\ref{KPZ_scaling}), it follows that,
if the dynamics of the phase belongs to the KPZ universality class,  the first-order coherence of the condensate is expected to decay following stretched exponentials in both space and time as
\begin{equation}
    g^{(1)}(\delta \boldsymbol{r}, \delta t) \sim \left\{\begin{array}{l l}
   \exp\left(-\frac{A}{2}\delta t^{2\beta}\right) \qquad&  {\rm for} \,\,|\delta \boldsymbol{r}|=0\\
\exp\left(-\frac{B}{2}|\delta \boldsymbol{r}|^{2\chi}\right) & {\rm for} \,\, \delta t=0
\end{array}\right. .
\end{equation}

Let us emphasize that, however, for a finite-size system, at long times,  the decay of the first-order coherence departs from the KPZ prediction, and instead  follows a simple exponential \cite{Keeling2010}, \textit{i.e.} 
\begin{equation}
g^{(1)}(\delta \boldsymbol{r}=\boldsymbol{0}, \delta t)\sim e^{-\delta t/\tau}\,
\label{eq:st}
\end{equation}
where $\tau$ is the associated relaxation time. This property is very general, as it holds in any dimension,  it was originally predicted in laser systems  and known as the
Schawlow--Townes regime
\cite{Schawlow1958, Fabre2010}. In the context of polaritons, the emergence of such long-time exponential decay has been evidenced in numerical simulations 
 \cite{Amelio2024}. The onset of the exponential Schawlow--Townes decay depends on the system size, and was observed to occur at smaller $\delta t$ for smaller sizes.
 Besides, for a finite size system, the KPZ prediction is also expected to be superseded at large $|\delta {\bf r}|$ by finite size effects, since the spatial decay is then affected  by the boundaries of the system, which prevent fluctuations from ever-growing (as predicted by the KPZ scaling). Thus, the KPZ universal scalings can only be found over a space-time window before these finite-size effects set in.

\subsection{\label{sec:theory_norm}Normalization  for a reliable collapse}

Note that in experiments, the  short-time and short-distance behavior is not universal, such that a transient precedes the onset of the KPZ behavior. This transient originates from the short-range coherence of the photoluminescence emitted by the high-energy, non-condensed polariton fraction. Thus, if one extrapolates the fit of the first-order coherence according to the model \eqref{KPZ_scaling}, obtained within the relevant KPZ space-time window, the extrapolated value at $(\delta {\bf r}=0,\delta t=0)$ will not be unity, {\it i.e.}  $g_{\rm fit}^{(1)}(\delta {\bf r}=0,\delta t=0)={1}/{\kappa}\neq 1$, whereas it should be normalized in order to observe to collapse of the data onto the universal scaling curve (see below). As shown in Ref.~\cite{Fontaine2022}, this can be accounted for by introducing the normalization factor $\kappa$, such that  $\kappa g_{\rm fit}^{(1)}(\delta {\bf r}=0,\delta t=0)={1}$. Multiplying by $\kappa$ the experimental $g_{\rm exp}^{(1)}(\delta \boldsymbol{r}, \delta t)$, one obtains the normalized $g_{\rm norm}^{(1)}(\delta {\bf r},\delta t)$:
\begin{equation}
g_{\rm norm}^{(1)}(\delta {\bf r},\delta t) = \kappa \, g_{\rm exp}^{(1)}(\delta \boldsymbol{r}, \delta t) \, ,
\end{equation}
Using this definition, one finds:
\begin{equation}
-2  \ln\left( |g_{\rm exp}^{(1)}(\delta t, \delta \boldsymbol{r})|\right)=
2 \ln(\kappa) -2\ln\left(|g_{\rm norm}^{(1)}(\delta t, \delta \boldsymbol{r})|\right)\,. 
\end{equation}
Therefore, if $-2\ln(|g_{\rm norm}^{(1)}(\delta t, \delta \boldsymbol{r})|)$ follows the KPZ scaling form \eqref{KPZ_scaling},  then plotting the experimental data $-2  \ln( |g_{\rm exp}^{(1)}(\delta t, \delta \boldsymbol{r})|)/t^{2\beta}$ as a function of the scaling variable $|\delta\boldsymbol{r}|/\delta t^{1/z}$ will lead to 
\begin{equation}
 -2\ln(|g_{\rm exp}^{(1)}(\delta t, \delta r)|)/\delta t^{2\beta} = C_0 \, F(C_1|\delta\boldsymbol{r}|/\delta t^{1/z}))+2 \, \ln(\kappa) / \delta t^{2\beta}    \,,
\end{equation}
 which will fail to collapse onto the universal scaling function $F$ because of the residual $1/\delta t^{2\beta}$ dependence. This holds not only when $\beta$ and $\chi$ take the KPZ values, but also for any values.
Conversely, as we show in the following, the term $2  \ln(\kappa) / t^{2\beta}$ can also have a very misleading influence, as it can seemingly improve the quality of a collapse, while being just an artifact of inappropriate normalization.

\section{Space and time decay of the first-order coherence in the experimental data}
\label{sec:scaling} 

\begin{figure}
\includegraphics[width=\textwidth]{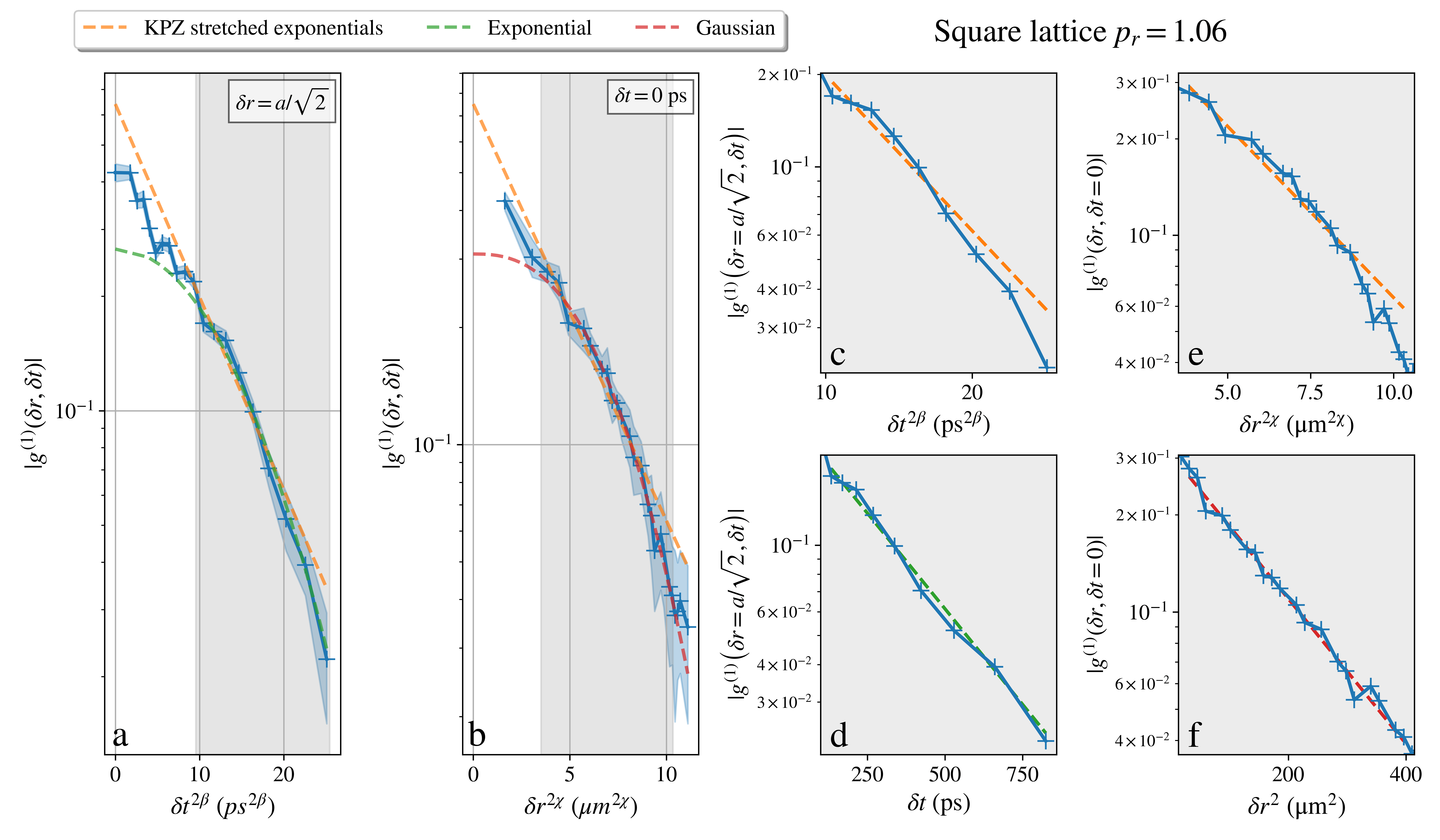}
\caption{\textbf{Experimental data for the first-order correlation used in Fig.~2\textbf{C} and Fig.~3\textbf{A} of Ref.~\cite{Widmann2026}, obtained on a square lattice and for $\bf p_r=1.06$.}  {\bf a}~Semi-logarithmical plot of the measured $|g^1(|\delta \boldsymbol{r}| = a/\sqrt{2},\delta t) |$ as a function of $\delta t^{2 \beta}$ with $\beta = 0.24$ (blue crosses). The orange (green) dashed lines show stretched exponential (Schawlow--Townes) scalings. \textbf{b.}~Semi-logarithmical plot of the measured $|g^1(\delta \boldsymbol{r},\delta t = 0) |$ as a function of $|\delta \boldsymbol{r}|^{2 \chi}$ with $\chi = 0.39$. The orange (red) dashed lines show stretched exponential (Gaussian) scalings. \textbf{c} to \textbf{f} Same data zoomed in the space-time windows corresponding to the gray shades displayed as a function of \textbf{c} $\delta t^{2\beta}$, \textbf{d} $\delta t$, \textbf{e} $\delta r^{2\chi}$, \textbf{f} $\delta r^2$, highlighting the curvature of the data with respect to the KPZ scalings (\textbf{c} and \textbf{e}) as opposed to exponential-Gaussian scalings (\textbf{d} and \textbf{f}).}
\label{fig:scalings-1}
\end{figure}

\begin{figure}
\includegraphics[width=\textwidth]{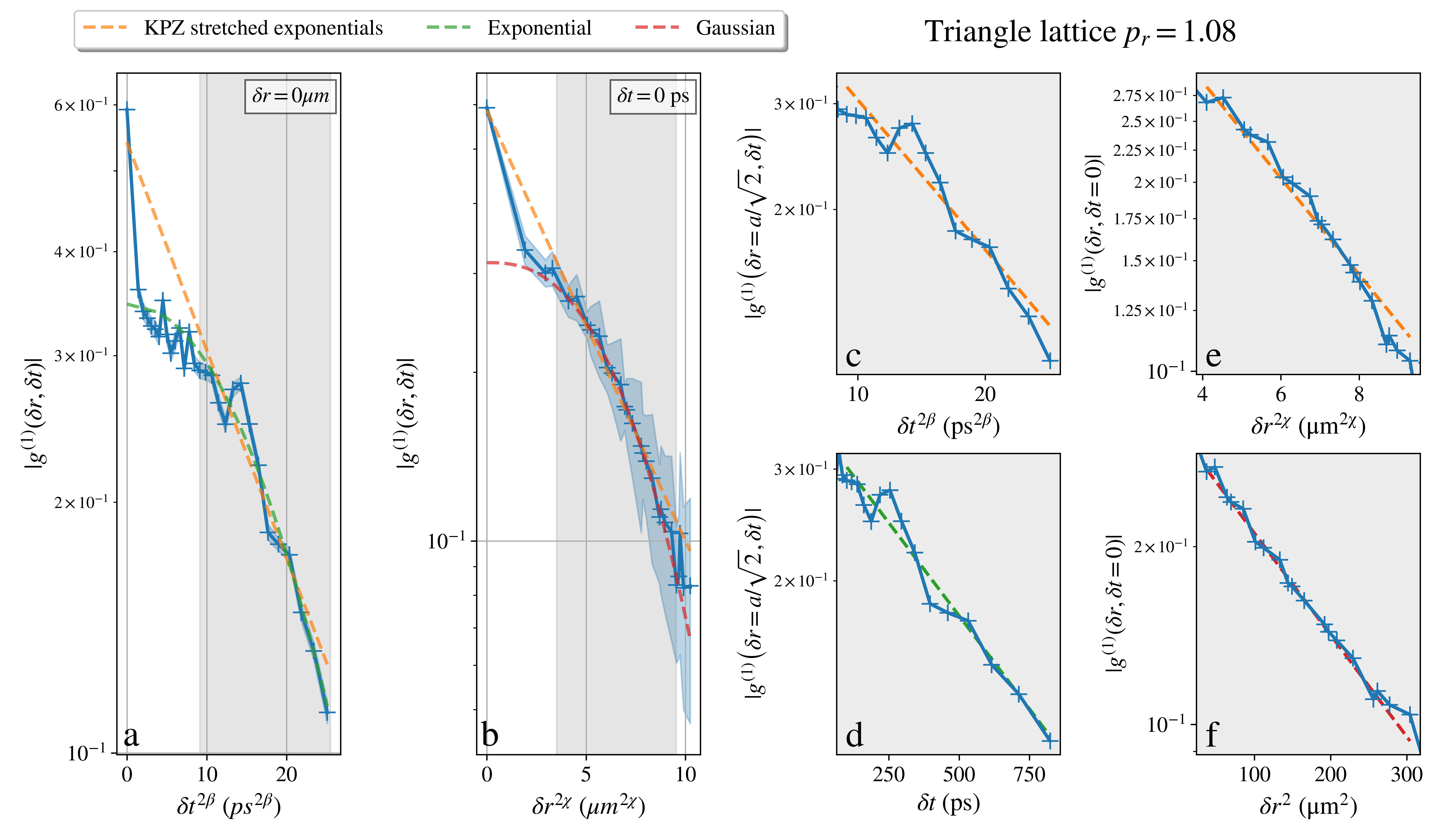}
\caption{\textbf{Experimental data for the first-order correlation used in Fig.~3\textbf{b} of Ref.~\cite{Widmann2026}, obtained on a triangular lattice and for $\bf p_r=1.07$}. {\bf a}~Semi-logarithmical plot of the measured $|g^1(|\delta \boldsymbol{r}| = a/\sqrt{2},\delta t) |$ as a function of $\delta t^{2 \beta}$ with $\beta = 0.24$ (blue crosses). The orange (green) dashed lines show stretched exponential (Schawlow--Townes) scalings. \textbf{b.}~Semi-logarithmical plot of the measured $|g^1(\delta \boldsymbol{r},\delta t = 0) |$ as a function of $|\delta \boldsymbol{r}|^{2 \chi}$ with $\chi = 0.39$. The orange (red) dashed lines show stretched exponential (Gaussian) scalings. \textbf{c} to \textbf{f} Same data zoomed in the space-time windows corresponding to the gray shades displayed as a function of \textbf{c} $\delta t^{2\beta}$, \textbf{d} $\delta t$, \textbf{e} $\delta r^{2\chi}$, \textbf{f} $\delta r^2$, highlighting the curvature of the data with respect to the KPZ scalings (\textbf{c} and \textbf{e}) as opposed to exponential-Gaussian scalings (\textbf{d} and \textbf{f}).}
\label{fig:scalings-1_triangle}
\end{figure}

\noindent
Experimental data for the space-time decay of the first-order coherence, as reported in Fig.~3\textbf{A} of Ref.~\cite{Widmann2026},  obtained on the square lattice and for $p_r=P/P_{\rm th}=1.061$, are displayed in Fig.~\ref{fig:scalings-1}. Similarly, we display in Fig.~\ref{fig:scalings-1_triangle} the data that Ref.~\cite{Widmann2026} reports on the triangular lattice, for the closest available pumping rate to the value $p_r=1.079$ which is used in Fig.~3\textbf{B}, \textit{i.e.} for $p_r=1.077$ (data for $p_r=1.079$ not provided). In both figures, the orange dashed line on panels \textbf{a} and \textbf{b} represent stretched exponential fits obtained using the KPZ exponents, and performed within the gray-shaded space-time window that the authors used to produce the data collapse in their Fig.~3\textbf{A} and Fig.~3\textbf{B}  (see Sec.~\ref{sec:norm_experimental}). The first-order coherence is displayed on a vertical semilog scale as a function of $\delta t^{2\beta}$ and $\delta r^{2\chi}$ respectively. Thus, as for the orange fits, a KPZ decay should appear as experimental data aligning along a straight line. Instead, they appear rather curved within the gray shade. Moreover, we find that the first-order coherence is better fitted in time (resp. in space), and over a broader range, using the 
Schawlow--Townes (Eq.~\ref{eq:st}) (resp. Gaussian) decay.
 This holds true for both the square and triangular lattices.

\bigbreak
\noindent
This is further highlighted on the right panels of Fig.~\ref{fig:scalings-1} and Fig.~\ref{fig:scalings-1_triangle}, where the same experimental data within the same gray-shaded windows  are plotted as $\log|g^{(1)}|$ either as a function of a KPZ power-law growth on panels \textbf{c} and \textbf{d}, or as a function of a linear (resp. quadratic) growth in time on panels \textbf{e} (resp. in space on panels \textbf{f}).
A model-insensitive way to highlight a stretched exponential decay and identify spatial and temporal critical exponents in the data  is to show $-\log|g^{(1)}|$  as a function of chosen powers of $|\delta {\bf r}|$  and $\delta t$. The data will follow a linear behavior only if the relevant power-law exponent is chosen, while they will appear as curved if the exponents are chosen differently.
It is clear from panels \textbf{c} to \textbf{f} that experimental data of Fig.~\ref{fig:scalings-1} and Fig.~\ref{fig:scalings-1_triangle} are better captured by an exponential (resp. Gaussian) decay of the first-order coherence in time (resp. in space), as anticipated from  panels \textbf{a} and \textbf{b}. 

 In order to quantitatively support this observation, we provide in Table~\ref{table} the weighted least square sums (WLS) between the experimental data and the different fitting models, within the space-time windows considered in Ref.~\cite{Widmann2026}.  For both spatial and temporal coherence
and for both lattice geometries, the values of the WLS are systematically and significantly larger for the KPZ model, indicating that the exponential-Gaussian one provides a more accurate description of the data. 

\begin{table}[h]
\centering
\begin{tabular}{| l | c c | c c |}
\hline\hline
 & \multicolumn{2}{ c |}{ WLS for temporal decay } & \multicolumn{2}{ c |}{ WLS for spatial decay } \\
 & \qquad   KPZ \qquad  \qquad  & \qquad Schawlow--Townes\qquad  &   \qquad KPZ \qquad  \qquad   & \qquad Gaussian \qquad\\
\hline
Square lattice ($p_r = 1.061$) & 12.97 & 4.07 & 10.83 & 4.78 \\
Triangular lattice ($p_r = 1.077$) & 204.6 & 125.5 & 2.06 & 0.66 \\
\hline\hline
\end{tabular}
\caption{Weighted least square (WLS) residuals between the data for the experimental first-order coherence and different models: KPZ stretched exponential in space and time, Schawlow--Townes temporal decay and spatial Gaussian decay, for both lattice geometries.}
\label{table}
\end{table}

\begin{figure}
\includegraphics[width=\textwidth]{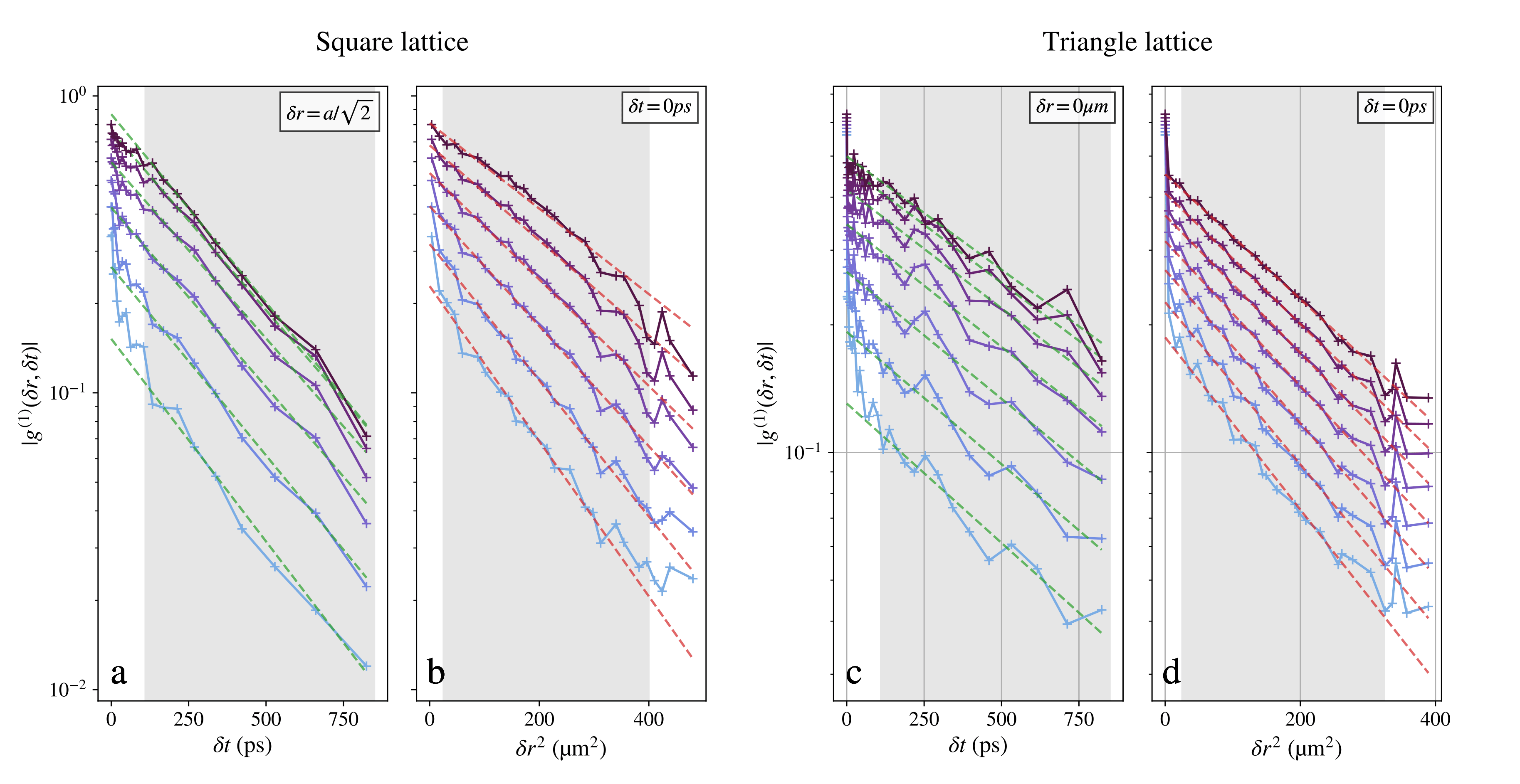}
\caption{\textbf{Experimental data for the first-order correlation used in Fig.~2\textbf{C} and Fig.~3 of Ref.~\cite{Widmann2026}.}  \textbf{a} and  \textbf{b}: data for the  square lattice and for pumping rates $p_r =\{1.042,1.061,1.079,1.094, 1.11, 1.12\}$ (from light blue to dark violet) as a function of \textbf{a} $\delta t$ for $\delta r = a/\sqrt{2}$ and \textbf{b} $\delta r^2$ for $\delta t = 0$;  \textbf{c} and \textbf{d}: data for  the triangular lattice and for  pumping rates  $p_r=\{1.060, 1.066, 1.072, 1.077, 1.081, 1.085, 1.088\}$  (from light blue to dark violet) as a function of \textbf{c} $\delta t$ for $\delta r = a/\sqrt{2}$ and \textbf{d} $\delta r^2$ for $\delta t = 0$;  Green (resp. red) dashed lines represent exponential (resp. Gaussian) fits.}
\label{fig:scalings-all}
\end{figure}

\bigbreak
\noindent
The same analysis is performed for all excitation powers reported in Ref.~\cite{Widmann2026} on the square lattice (Fig.~2\textbf{C} and Fig.~3\textbf{C}) and on the triangular lattice (Fig.~3\textbf{D}). The result  is displayed in Fig.~\ref{fig:scalings-all}. We find that, for every pumping powers, the first-order coherence is systematically better captured by an exponential decay (resp. Gaussian) in time (resp. in space).

\section{\label{sec:collapse}Collapse of experimental data}

\begin{figure}
\includegraphics[width=1.08\textwidth]{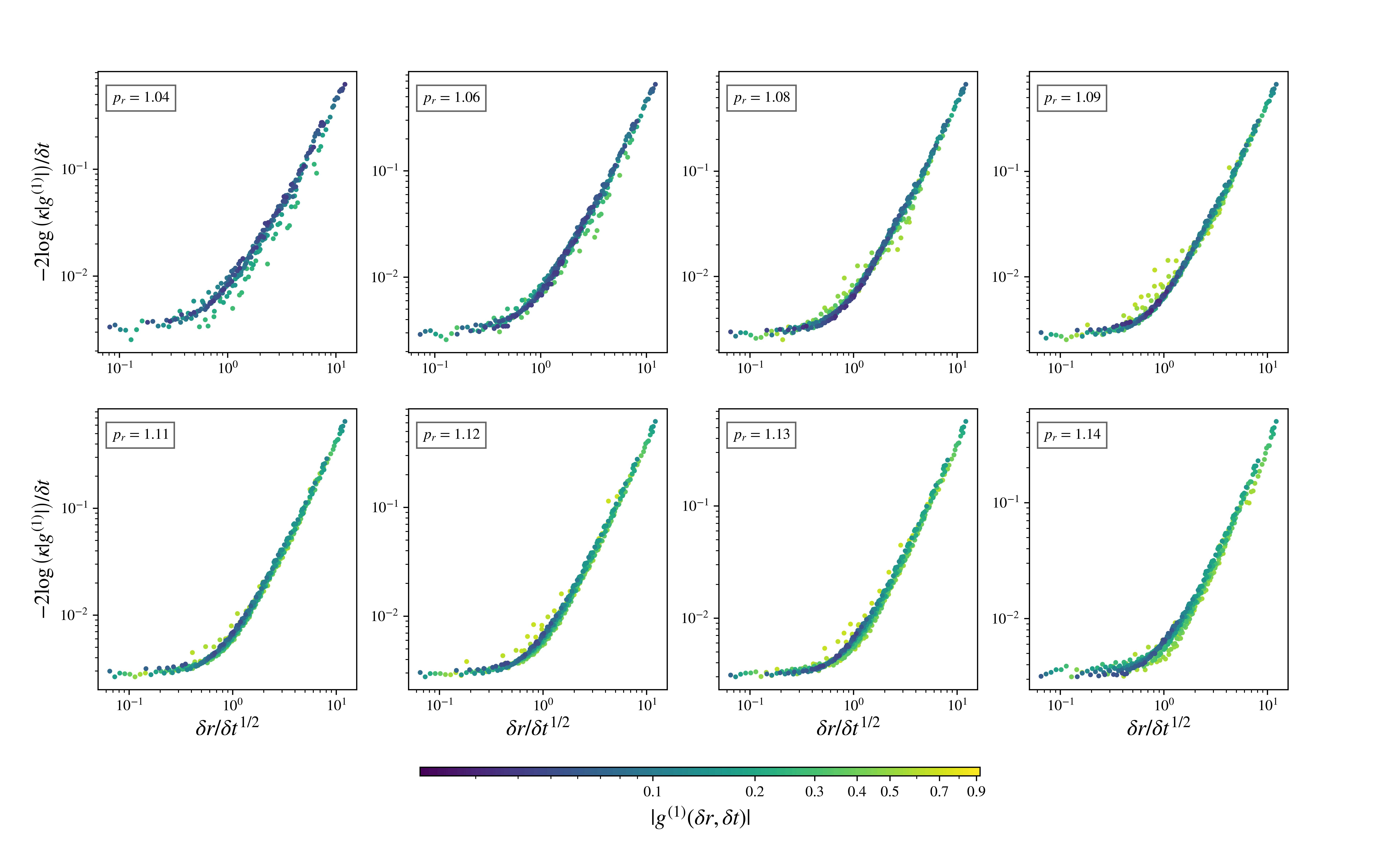}
\caption{\textbf{Data collapse of the full space-time map of the first-order correlation function measured on the square lattice using $\bf \beta = 0.5$ and $\bf \chi = 1$}  and for increasing pumping rates from top left to bottom right. The colorscale indicates coherence, from yellow (high coherence) to blue (low coherence).
}
\label{fig:collapse-all-carre}
\end{figure}

\noindent
In this section, we show that the space-time map of the first-order coherence measured on the square lattice can be collapsed onto a universal function for every pumping rates if one sets $\beta=0.5$ and $\chi=1$. Choosing these exponents reproduce an exponential decay in time and a Gaussian decay in space. The different collapses are displayed in Fig.~\ref{fig:collapse-all-carre}, where $g^{(1)}$ maps have been normalized as explained in Sec.~\ref{sec:theory_norm}. The results for the triangular lattice are shown in  Appendix~\ref{app}. 

\noindent
The quality of the data collapse shown in Fig.~\ref{fig:collapse-all-carre} and  Appendix~\ref{app} is to be contrasted with collapses reported in Fig.~3.\textbf{A} (square lattice) and Fig.~3.\textbf{B} (triangular lattice) of Ref.~\cite{Widmann2026} using KPZ exponents (see also their Supplementary Materials, and  Appendix~\ref{app} where they are reproduced). In these figures, the KPZ collapse seemingly works for $p_r=1.061$ (square lattice, left panel in Fig.~\ref{fig:scalings-all_square}) and $(p=1.077)$ (triangular lattice, left panel in Fig.~\ref{fig:scalings-all_triangle}) only, while datapoints are very scattered for higher pumping rates. This point is further argued in Sec.~\ref{sec:norm_experimental} hereafter.

\begin{figure}
\includegraphics[width=1.08\textwidth]{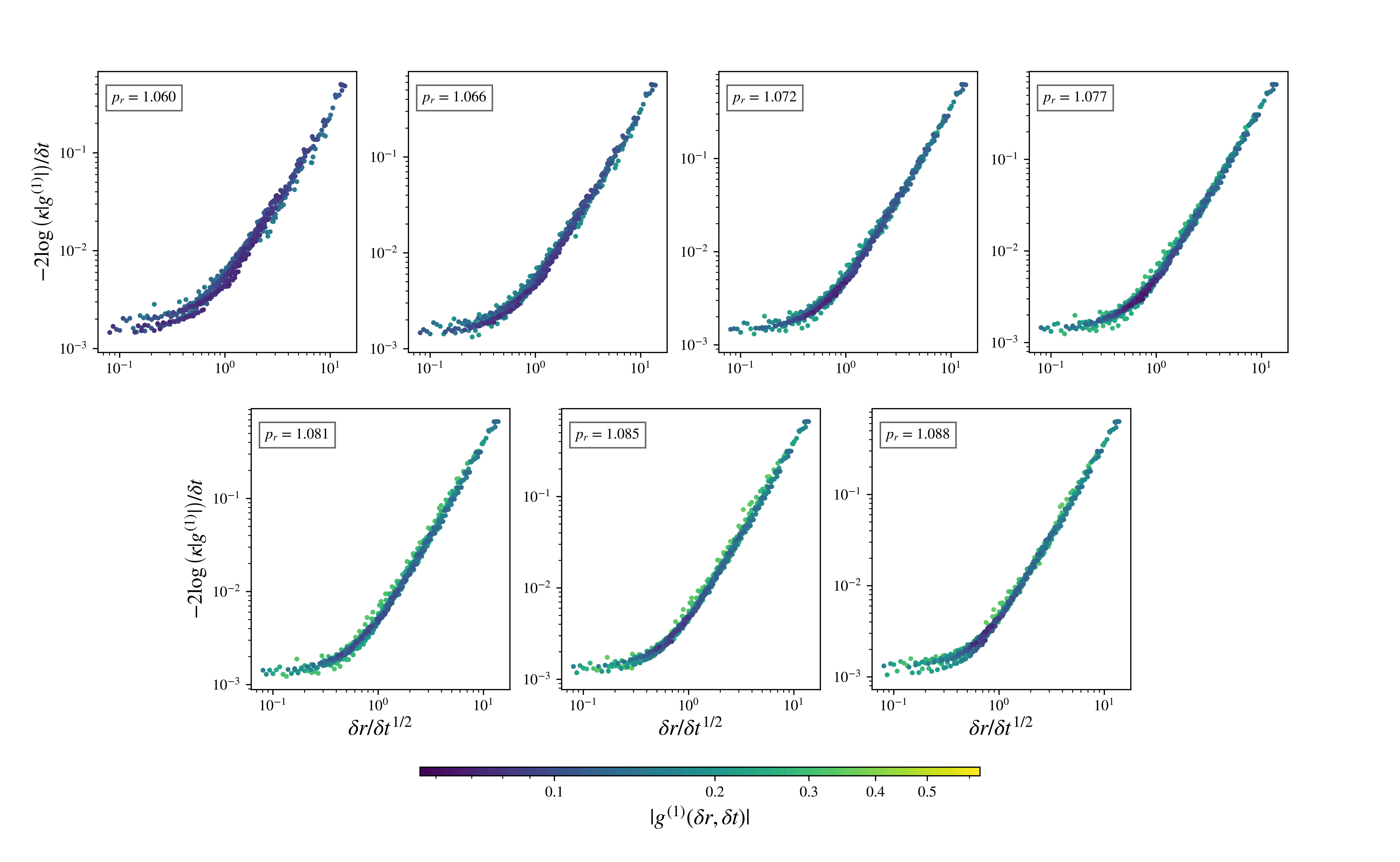}
\caption{\textbf{Data collapse of the full space-time map of the first-order correlation function measured on the triangular lattice using $\bf \beta = 0.5$ and $\bf \chi = 1$}  and for increasing pumping rates from top left to bottom right. The colorscale indicates coherence, from yellow (high coherence) to blue (low coherence).}
\label{fig:collapse-all-triangles}
\end{figure}

\noindent

\section{\label{sec:norm}Importance of normalization}

\noindent
In this section, we show the influence of the normalization of the data on the quality of a collapse. We first demonstrate this on  idealized $g^{(1)}$ maps generated from  analytic formula, before turning to the analysis of the experimental data.

\subsection{\label{sec:norm_analytical}Analytically-generated data}

\begin{figure}
    \centering
    \includegraphics[width=0.8\linewidth]{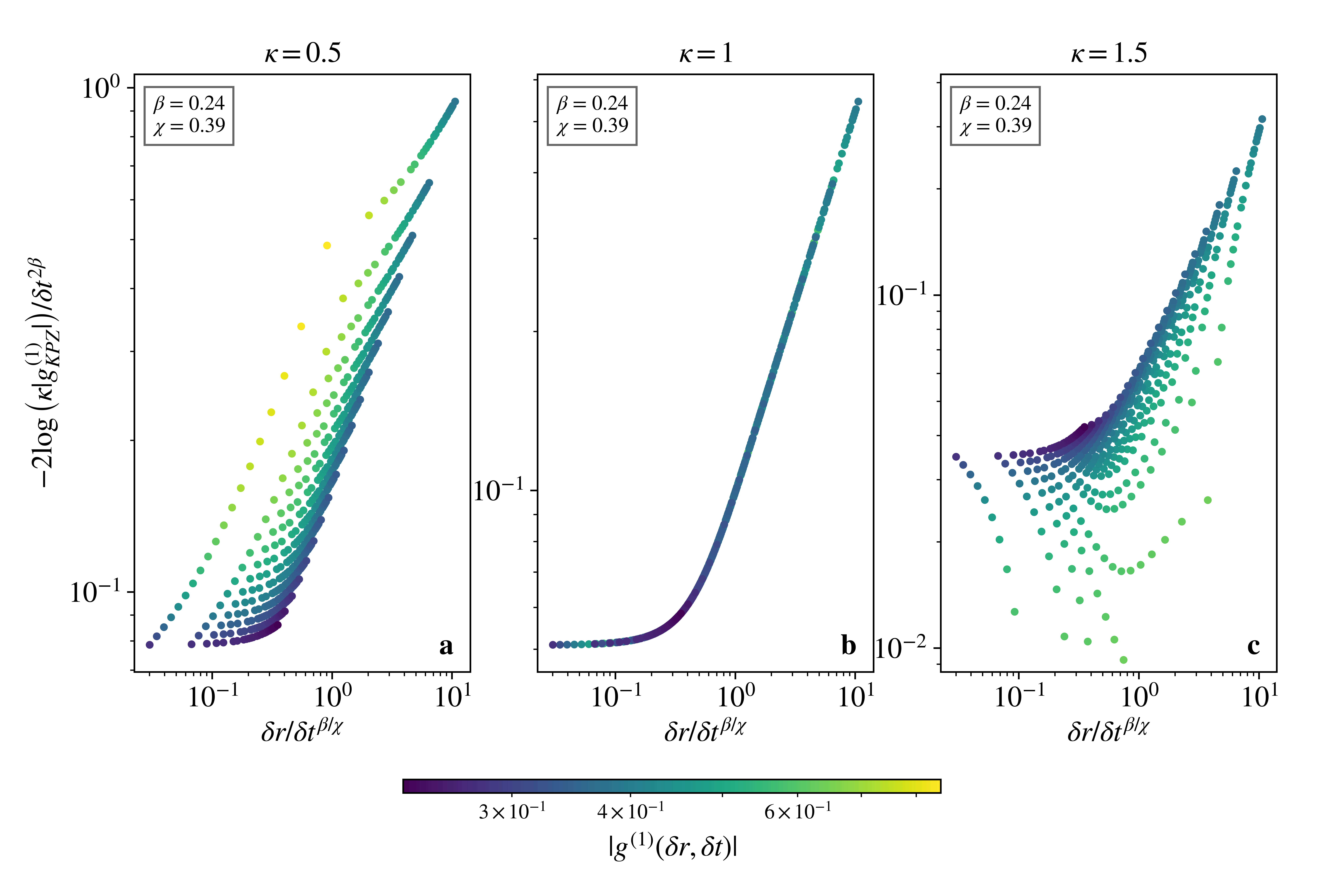}
    \caption{\textbf{Data collapse obtained from the dataset of $\bf g_{\rm KPZ}^{(1)}$ generated according to Eq.~\eqref{eq:mapg1}, \textit{i.e.} following the (exponential of the) exact KPZ scaling form} for three different normalization factors \textbf{a} $\kappa=0.5$, \textbf{b} $\kappa=1$, \textbf{c} $\kappa=1.5$. Only the appropriate normalization $\kappa=1$ yields the expected data collapse onto a one-dimensional curve.}
    \label{fig:KPZ_theor_collapse}
\end{figure}

\noindent
We generate a dataset for $g^{(1)}(\delta{\bf r},\delta t)$ following  the exact KPZ universal behavior, {\it i.e.} using the analytical expression
\begin{equation}
g_{\rm KPZ}^{(1)}(\delta{\bf r},\delta t) = \exp\left[-\delta t^{2\beta}\frac{C_0}{2}F\left(y_0 \frac{|\delta {\bf r}|}{\delta t^{1/z}}\right) \right]
\label{eq:mapg1}
\end{equation}
with $\beta = 0.24$, $z=1.62$, $C_0 =0.005$, and $y_0=0.15$, and $F$ the KPZ universal scaling function. The proper normalization factor for this dataset is  $\kappa=1$.
A data collapse on the scaling function $F$ is thus expected if one represents $-2\ln|g_{\rm KPZ}^{(1)}|/\delta t^{2\beta}$ as a function of $|\delta {\bf r}|/\delta t^{1/z}$. We show in Fig.~\ref{fig:KPZ_theor_collapse} the result obtained setting different values of the normalization $\kappa$, {\it i.e.}
 we display $-2\ln|\kappa g_{\rm KPZ}^{(1)}|/\delta t^{2\beta}$, using the generated dataset $g_{\rm KPZ}^{(1)}$ for   different  $\kappa$. One observes  that the expected collapse is only obtained with the proper normalization $\kappa=1$ (Fig.~\ref{fig:KPZ_theor_collapse}.\textbf{b}). If the value of $\kappa$ is changed, the collapse is completely spoiled (Fig.~\ref{fig:KPZ_theor_collapse}.\textbf{a} and \textbf{c}). This illustrates that an inappropriate  normalization has a dramatic effect on the collapse.

\begin{figure}
    \centering
    \includegraphics[width=0.8\linewidth]{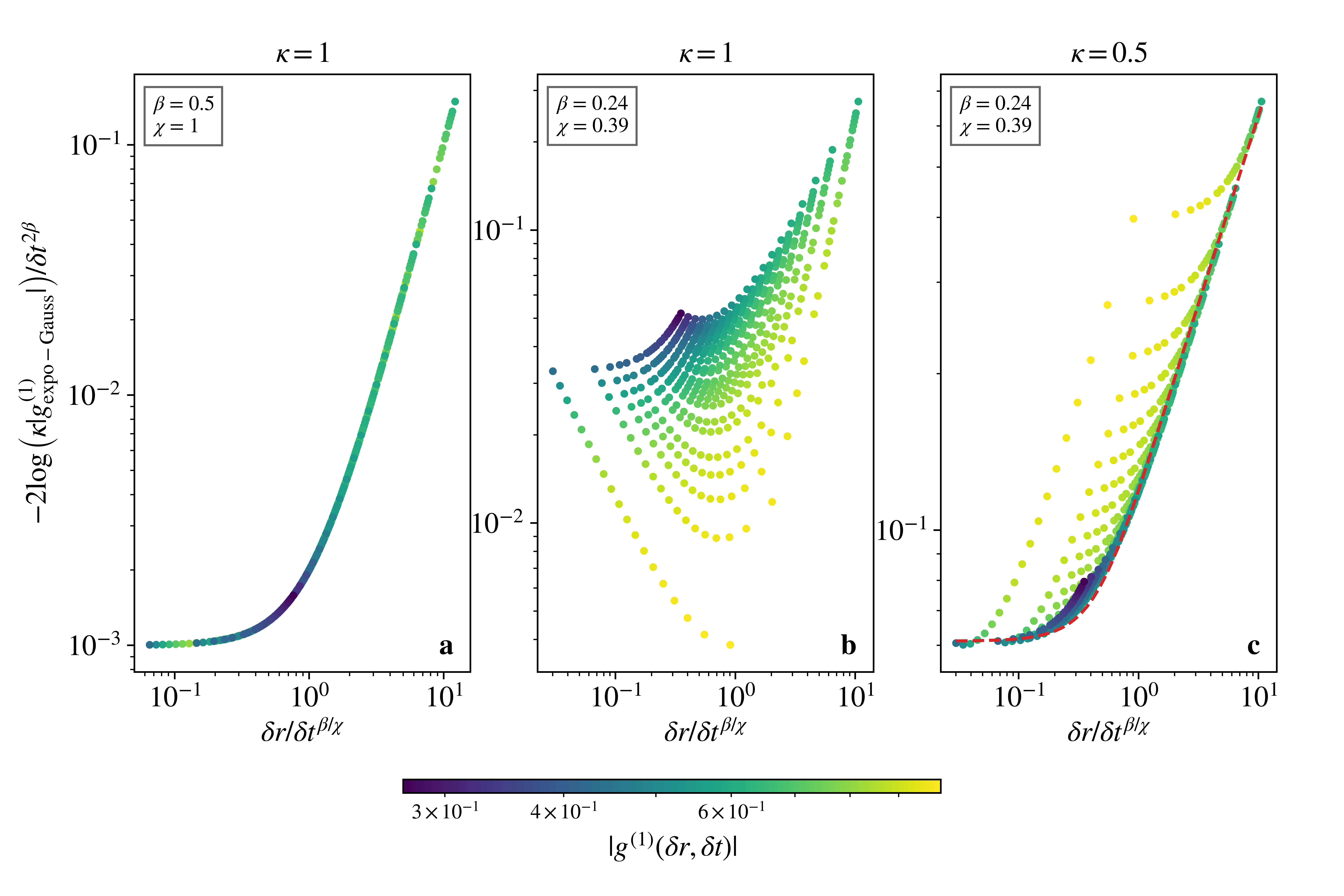}
    \caption{\textbf{Data collapse obtained from the dataset of $\bf g_{\rm expo-Gauss}^{(1)}$ generated according to Eq.~\eqref{eq:mapg1-exp}, \textit{i.e.} following an exponential decay in time, and a Gaussian decay in space} in three different representations \textbf{a} using the correct exponents $\beta=0.5$ and $\chi=1$ corresponding to Eq.~\eqref{eq:mapg1-exp} and appropriate normalization $\kappa=1$, \textbf{b} using the (incorrect) KPZ exponents $\beta=0.24$ and $\chi=0.39$  and appropriate normalization $\kappa=1$, \textbf{c} using the (incorrect) KPZ exponents $\beta=0.24$ and $\chi=0.39$  and the inappropriate normalization $\kappa=0.5$. The red dashed curve is the KPZ universal scaling function \cite{Kloss2012}, showing how misleading a collapse with inappropriate normalization can be.}
    \label{fig:Exp_Gauss_theor_collapse}
\end{figure}

\bigbreak
\noindent
To further illustrate the misleading effect which can originate from an  inappropriate  normalization, we generate a second dataset for $g^{(1)}(\delta{\bf r},\delta t)$ following an exponential decay in time, and a Gaussian decay in space, {\it i.e.} using the analytical expression
\begin{equation}
g_{\rm expo-Gauss}^{(1)}(\delta{\bf r},\delta t) = \exp\left[-\frac{\delta t}{T}\left(1+ \frac{|\delta {\bf r}|^2}{R^2\delta t}\right) \right]\,,
\label{eq:mapg1-exp}
\end{equation}
with $T=10^{-3} \mathrm{ps}$, and $R=10^{-3} \mathrm{\mu m}$. The appropriate normalization is again $\kappa=1$.
In this case, a data collapse is expected if one represents $-2\ln| g_{\rm expo-Gaus}^{(1)}|/\delta t$ as a function of the scaling variable $|\delta {\bf r}|/\delta t^{1/2}$, onto the scaling function which is simply the parabola $F(x) = 1/T + (x/R)^2$. We show in Fig.~\ref{fig:Exp_Gauss_theor_collapse}.\textbf{a} that a perfect collapse is indeed obtained as expected in this representation. 
However, we also show in Fig.~\ref{fig:Exp_Gauss_theor_collapse}.\textbf{b} and \textbf{c} the same data but represented using the KPZ scaling exponents instead of the correct exponents, which are $\beta=0.5$ (exponential) and $\chi=1$ (Gaussian) for this data set. Specifically we display $-2\ln| \kappa g_{\rm expo-Gaus}^{(1)}|/\delta t^{2\beta}$ with $\beta=0.24$ as a function of $|\delta {\bf r}|/\delta t^{1/z}$ with $z=\chi/\beta=1.62$, for two values of the normalization $\kappa$. With the appropriate normalization (Fig.~\ref{fig:Exp_Gauss_theor_collapse}.\textbf{b}), the data do not collapse at all on a single one-dimensional curve, as expected since it is not the adapted representation for these data. This picture is reminiscent of Figs.~S2\textbf{B} to \textbf{D} in the Supplementary Material of Ref.~\cite{Widmann2026}. However, using a modified (inappropriate) normalization (Fig.~\ref{fig:Exp_Gauss_theor_collapse}.\textbf{c}), the situation is seemingly improved, as two asymptotes reproducing the KPZ scaling function (red dashed line) seem to appear. This picture is very reminiscent of Fig.~3\textbf{B} of Ref.~\cite{Widmann2026}, reported in Fig.~\ref{fig:collapse-1-triangle}.\textbf{a} hereafter (see also Figs.~S2\textbf{E} to \textbf{H} of the Supplementary Material of Ref.~\cite{Widmann2026}). This illustrates the influence of the normalization, which can in certain cases, when inappropriately determined, hide the true features of a dataset and mislead for alien ones.
This analysis of the effect of the normalization is applied to the experimental data in the next section.

\subsection{\label{sec:norm_experimental}Experimental data}

\noindent
In the light of the preceding sections, we re-analyze the data collapse of Ref.~\cite{Widmann2026} provided in the main article for the square and triangular lattice (respectively Figs.\ref{fig:collapse-1-square} and \ref{fig:collapse-1-triangle}). We show the data collapse using two representations: 

(i) the KPZ one used in the Ref.~\cite{Widmann2026}, which consists in plotting $-2\ln| \kappa g_{\rm exp}^{(1)}|/\delta t^{2\beta}$ as a function of $|\delta {\bf r}|/\delta t^{1/z}$ with the KPZ exponents $\beta=0.24$ and  $z=1.62$; 

(ii) the exponential-Gaussian one as used in Sec.~\ref{sec:collapse}, which consists in plotting $-2\ln| \kappa g_{\rm exp}^{(1)}|/\delta t^{2\beta}$ as a function of $|\delta {\bf r}|/\delta t^{1/z}$ with the exponents $\beta=0.5$ (exponential) and  $z=2$ (Gaussian). 

\bigbreak
\noindent
Let us emphasize that throughout this analysis, we use the same fixed space-time windows as in the previous section. This is different from Ref.~\cite{Widmann2026}, where a further selection is performed within these windows, eliminating certain data points. Here, we use all the data points available. 
For both lattice geometries, panel \textbf{c} shows the very satisfactory collapse obtained with representation (ii) already demonstrated in  Sec.~\ref{sec:collapse}, and without the need to discard data points. Panels \textbf{a} and \textbf{b} show the results obtained with representation (i): \textbf{a} without normalization ($\kappa=1$) which reproduces Figs.~3\textbf{A} and \textbf{B} in the main text of Ref.~\cite{Widmann2026} ; \textbf{b} with the appropriate normalization. It is clear that the (properly normalized) data do not collapse well in the KPZ representation, in accordance with the scaling analysis presented in  Sec.~\ref{sec:scaling}. However, the lack of normalization gives the misleading appearance of a good collapse.
The same analysis for all data available from Ref \cite{Widmann2026}, which correspond to different pump powers and lattice geometries,  is provided in the Appendix~\ref{app}. The conclusions are similar. The main outcome of this systematic analysis is hence to show the robustness of the exponential-Gaussian modeling, since it provides in all cases a very satisfactory collapse. 

\begin{figure}
    \centering
    \includegraphics[width=\linewidth]{collapses_square_compressed.png}
    \caption{\textbf{Data collapse for the space-time map of $\bf g^{(1)}(\delta \boldsymbol{r}, \delta t)$ measured on a square lattice and used in Fig.~3\textbf{A} of Ref.~\cite{Widmann2026}} obtained  {\bf b} using the KPZ exponents and without normalization, {\bf c} using the KPZ exponents and with normalization, \textbf{d} using $\beta = 0.5$, $\chi=1$ and with normalization. The orange (green) star on panel \textbf{a} shows the extrapolated value at $\delta t =0$ of the stretched exponential (Shawlow-Townes) fit, within the KPZ window, of the temporal decay of $ |g^{(1)}(a/\sqrt{2}, \delta t)|$, which determines the normalization factor $1/\kappa$.}
    \label{fig:collapse-1-square}
\end{figure}

\begin{figure}
    \centering
    \includegraphics[width=\linewidth]{collapses_triangle_compressed.png}
    \caption{\textbf{Data collapse for the space-time map of $\bf g^{(1)}(\delta \boldsymbol{r}, \delta t)$ measured on a triangular lattice and used in Fig.~3\textbf{B} of Ref.~\cite{Widmann2026}} obtained  {\bf b} using the KPZ exponents and without normalization, {\bf c} using the KPZ exponents and with normalization, \textbf{d} using $\beta = 0.5$, $\chi=1$ and with normalization. The orange (green) star on panel \textbf{a} shows the extrapolated value at $\delta t =0$ of the stretched exponential (Shawlow-Townes) fit, within the KPZ window, of the temporal decay of $ |g^{(1)}(0, \delta t)|$, which determines the normalization factor $1/\kappa$.}
    \label{fig:collapse-1-triangle}
\end{figure}

\section{\label{sec:conclusion}Conclusion}

In this comment, we have re-analyzed the data presented in Ref. \cite{Widmann2026}. We have shown that the data for the first-order coherence is best described, for all pump powers and for both the square and the triangular lattices, by an exponential (Scahwlow-Townes) decay in time and a Gaussian decay in space over the space-time ranges analyzed in this paper. This invalidates the claim of the paper for evidence of KPZ universal scaling.  We have in particular clearly demonstrated how an inappropriate normalization can distort a data collapse, and in the present case lead to the misleading appearance of a reasonable collapse with the KPZ exponents. Instead, we have provided robust evidence for a very satisfactory collapse on an exponential-Gaussian model with the properly normalized data, since it holds for both lattice geometries and all pump powers available. The conclusions of 
Ref.~\cite{Widmann2026} 
are thus not justified, as KPZ universal scaling is not the best modeling of the data in the space-time windows analyzed in the paper.

\section{\label{app}Appendix: additional data analysis}

\begin{figure}
\includegraphics[width=\textwidth]{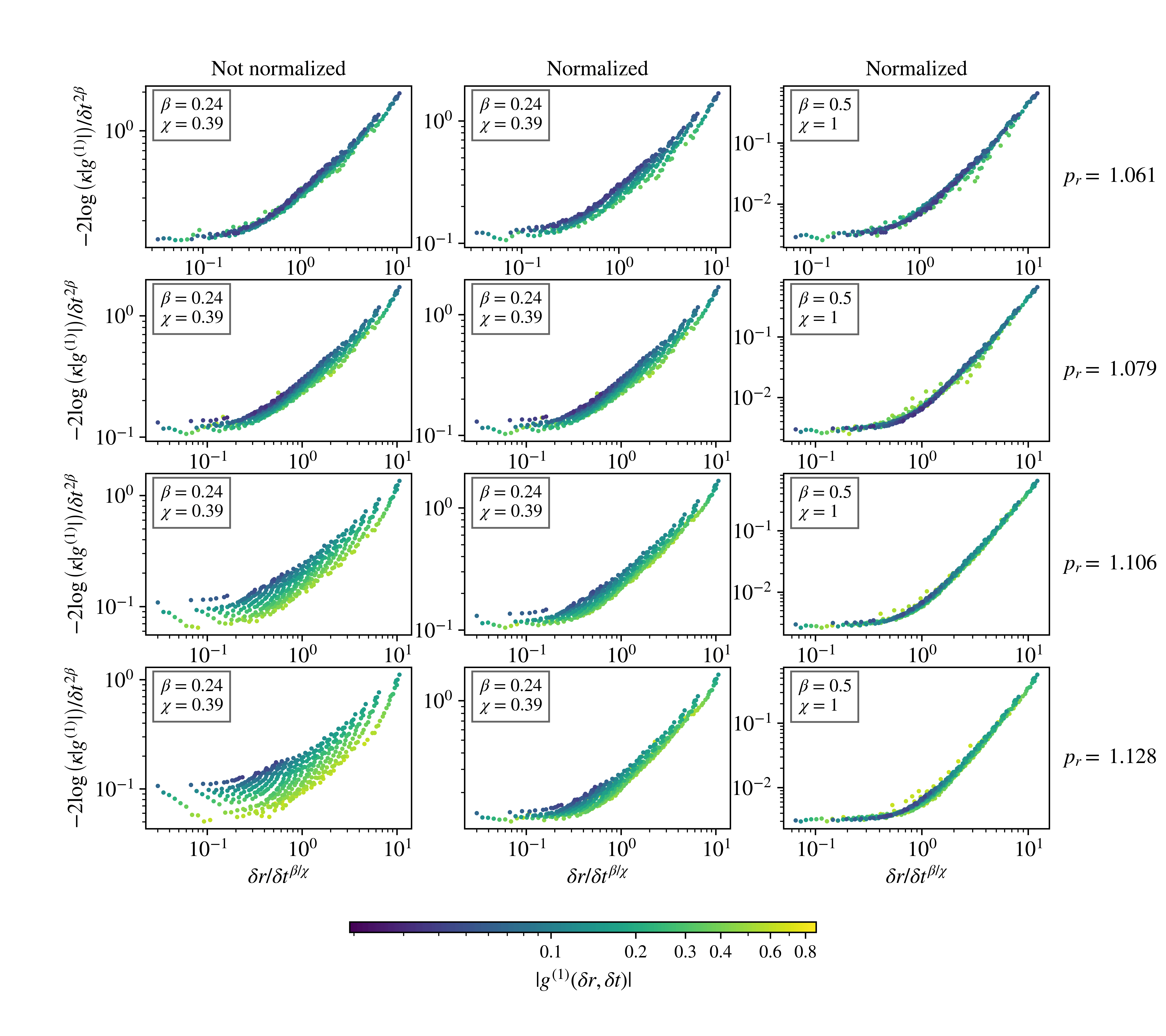}
\caption{\textbf{Data collapse for the space-time map of $\bf g^{(1)}(\delta \boldsymbol{r}, \delta t)$ measured on a square lattice in Ref.~\cite{Widmann2026}} for excitation powers $p_r=\{1.06, 1.08, 1.11, 1.13\}$, \textbf{left column} using the KPZ exponents and without normalization, \textbf{central column} using the KPZ exponents and with normalization, \textbf{right column} using $\beta = 0.5$, $\chi=1$ and with normalization.}
\label{fig:scalings-all_square}
\end{figure}

\begin{figure}
\includegraphics[width=\textwidth]{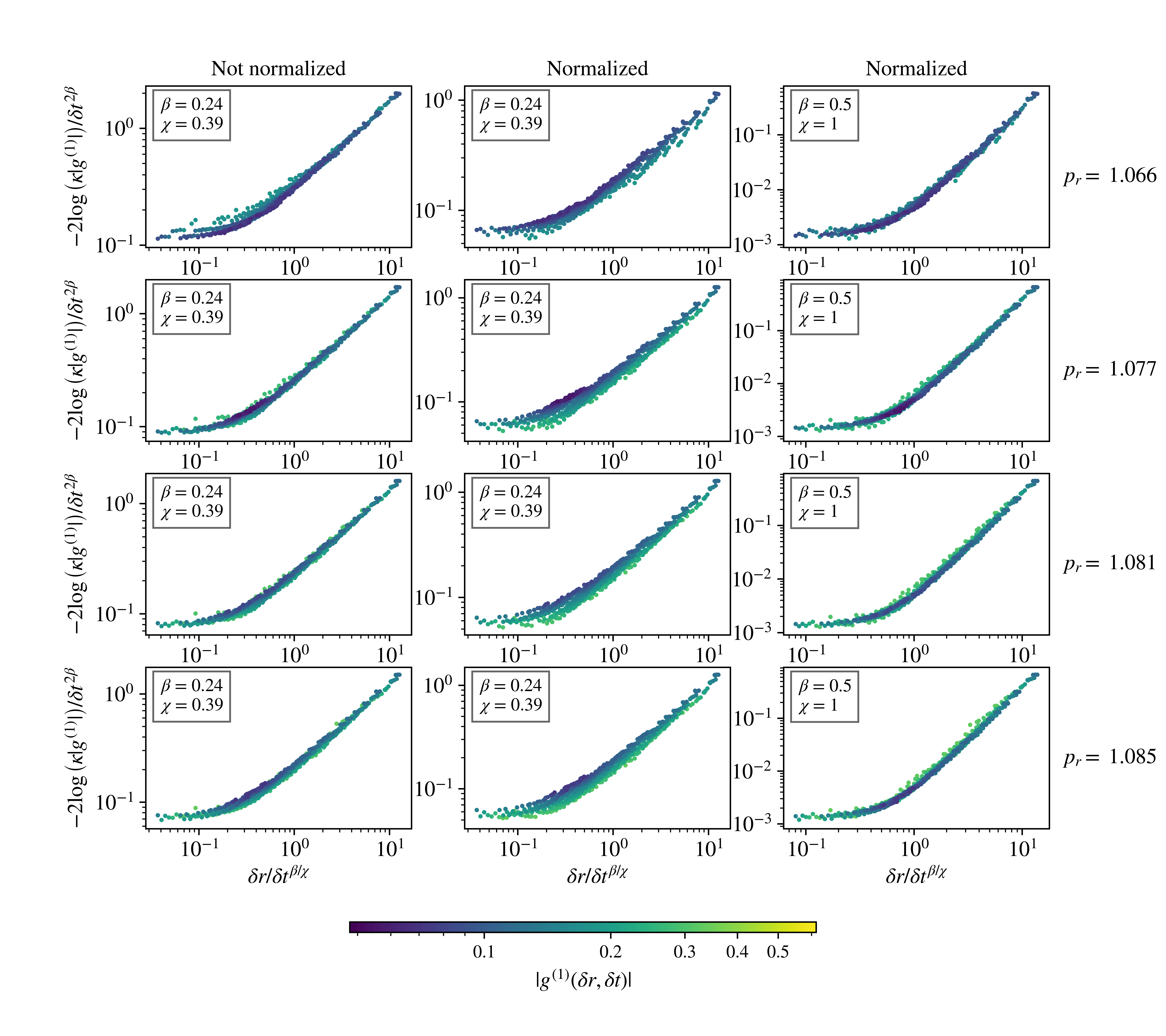}
\caption{\textbf{Data collapse for the space-time map of $\bf g^{(1)}(\delta \boldsymbol{r}, \delta t)$ measured on a triangular lattice in Ref.~\cite{Widmann2026}} for excitation powers $p_r=\{1.066, 1.077, 1.081, 1.085\}$, \textbf{left column} using the KPZ exponents and without normalization, \textbf{central column} using the KPZ exponents and with normalization, \textbf{right column} using $\beta = 0.5$, $\chi=1$ and with normalization.}
\label{fig:scalings-all_triangle}
\end{figure}

In this Appendix, we provide the same analysis of data collapses as performed in Sec. \ref{sec:norm_experimental} for all the available experimental data, \textit{i.e.} the square (resp. triangular) lattice geometry and different pump powers from $p_r=1.06$ to $p_r=1.13$ (resp. $p_r=1.066$ to $p_r=1.085$). We show in   in Fig. \ref{fig:scalings-all_square} (square lattice) and  Fig. \ref{fig:scalings-all_triangle} (triangular lattice) a representative subset of the available pump powers. For each of these pump powers, we display 

\medbreak
 -- in panel (a) and (b) the collapse obtained in representation (i) without normalization (panel a), which reproduces the figures shown in the Supplemental Material of Ref. \cite{Widmann2026}, and with appropriate normalization (panel b)

 \medbreak
-- in panel (c) the collapse obtained in representation (ii), with appropriate normalization.

\medbreak
\noindent
We have also analyzed  the other available pump powers (not shown), which  all yield very similar results. For both lattice geometry and all pump powers, the representation (ii), corresponding to exponential decay in time and Gaussian decay in space, always provides a very satisfactory collapse, which confirms the validity of this modelisation of the data. In contrast, the KPZ representation (i) do not yield a convincing collapse of the data, which demonstrates that these data in the considered ranges do not follow KPZ universal scalings.

\clearpage
%

\end{document}